\documentstyle[12pt]{article}

\newcommand {\beq}{\begin{equation}}
\newcommand {\eeq}{\end{equation}}
\newcommand {\beqa}{\begin{eqnarray}}
\newcommand {\eeqa}{\end{eqnarray}}         
\newcommand {\beqs}{\begin{eqnarray*}}
\newcommand {\eeqs}{\end{eqnarray*}}
\newcommand {\bds}{\begin{displaymath}}
\newcommand {\eds}{\end{displaymath}}
\newcommand {\n}{\nonumber\\}

\newcommand{\no}{\noindent}

\newcommand {\bebb}{\begin{thebibliography}}
\newcommand {\eebb}{\end{thebibliography}}      
\newcommand {\bbit}{\bibitem}

\newcommand{\cT}{{\cal T}}

\def\al{\alpha}
\def\bt{\beta}

\def\gm{\gamma}

\def\ph{\phi}

\def\ps{\psi}
\def\Ps{\Psi}

\def\sgm{\sigma}

\def\tht{\theta}

\def\tl{\tilde}  

\def\p{\partial}

\def\op{\oplus}

\def\psd{\psi ^{\dagger}}

\def\psb{\bar{\psi}}
\def\psbd{\bar{\psi} ^{\dagger}}
\def\psp{\psi ^{\prime}}
\def\pspd{{\psi ^{\prime}}^{\dagger}}

\def\journal#1&#2(#3){\unskip, \sl #1\ \bf #2 \rm(19#3) }
\def\andjournal#1&#2(#3){\sl #1~\bf #2 \rm (19#3) }

\def\npb#1#2#3{Nucl. Phys. {\bf B#1}, (#2) #3}

\def\plb#1#2#3{Phys. Lett. {\bf B#1}, (#2) #3}
\def\prl#1#2#3{Phys. Rev. Lett. {\bf #1}, (#2) #3}

\def\cmp#1#2#3{Commun. Math. Phys. {\bf #1}, (#2) #3}

\begin{document}

\begin{titlepage}

\begin{flushright}
\end{flushright}

\vskip 1cm

\begin{center}
{\Large\bf On $osp(2|2)$ Conformal Field Theories}

\vspace{1cm}

{\normalsize\bf
Xiang-Mao Ding $^{a,b}$ 
Mark D. Gould $^a$, \\
Courtney J. Mewton $^a$ and Yao-Zhong Zhang $^a$
}
\vskip.1in
{\em $^a$ Centre for Mathematical Physics, Division of Mathematics, \\
University of Queensland, Brisbane, Qld 4072, Australia}
\\
{\em $^b$ Institute of Applied Mathematics, Academy of Mathematics 
and System Sciences; Chinese Academy of Sciences, P.O.Box 2734, 
100080, China.}

\end{center}

\date{}


\vspace{2cm}

\begin{abstract}

We study the conformal field theories corresponding to
current superalgebras $osp(2|2)^{(1)}_k$ and $osp(2|2)^{(2)}_k$.
We construct the free field realizations, screen currents and primary
fields of these current superalgebras at general level $k$. All the results
for $osp(2|2)^{(2)}_k$ are new, and the results for the primary fields of
$osp(2|2)^{(1)}_k$  also seem to be new. Our results are expected to be
useful in the supersymmetric approach to Gaussian disordered systems
such as random bond Ising model and Dirac model.

\end{abstract}

\vspace{1cm}


\vspace{0.5cm}

\end{titlepage}

\setcounter{section}{0}
\setcounter{equation}{0}
\section{Introduction}

In recent years, disordered systems have attracted much
attention in both theoretical and condensed matter physics
communities \cite{Dot83,Lud90,Sha87,Car82,Lud94,Zir94,Cau96,Bas00}. 
In particular the application of the
supersymmetric method \cite{Efe83} to Gaussian disordered systems has
revealed that the relevant algebras are current superalgebras
with zero superdimension \cite{Mud96,Ber95,Maa97,Gad91,Gur99}. 
Such superalgebras have equal number
of bosonic and fermionic generators. This ensures that the
Virasoro algebra constructed from the super currents has
vanishing central charge: a necessary condition for the
description of disordered systems. The conformal field theory
derived from such a current superalgebra potentially 
contains primary fields with negative conformal dimensions so that the
theory is non-unitary. The non-unitarity makes the conformal field theory
non-trivial even though it has a vanishing central charge.

Our aim in this paper is to provide some algebraic backgrounds
which are expected to be useful in the study of random bond Ising
model and two-species Dirac model with a random $sl(2)-$gauge
potential. Namely, we investigate the conformal field theories
based on the current superalgebras $osp(2|2)^{(1)}$ and
$osp(2|2)^{(2)}$ at general level $k$. We derive the free
field representations and screen currents of these two
algebras. Primary fields corresponding to both typical and atypical
representations are constructed explicitly and their operator product
expansions (OPEs) with
currents are presented. In the case of $osp(2|2)^{(1)}$, there exists
an infinite family of negative dimensional primary operators, and for
the case of $osp(2|2)^{(2)}$ all primary fields have zero conformal
dimensions. All results for $osp(2|2)^{(2)}_k$ are
new, and the explicit results for the primary fields of
$osp(2|2)^{(1)}_k$ also seem to be new. As for the 
free field realizations and screen currents of $osp(2|2)^{(1)}_k$,
similar results have also been obtained in \cite{BOo,Ito,Bow96,Ras98},
though based on different approaches and conventions. We used a bit more
straightforward  approach by means of super coherent states. 
The free field realizations of the currents are needed in order
to find all representations (i.e. primary fields) of the current superalgebras.

This paper is organized as follows. In section 2, we set our
convention.  As is well-known, free field realization is a common 
approach used in both conformal field theories and representation 
theory of current algebras
\cite{Wak,FF3,BMP3,Kur,Fren}. So in section 3, we describe
our construction of the free field representations and screen
currents. In section 4, we construct the primary fields
corresponding to both typical and atypical representations of
the current superalgebras. We conclude in section 5.

\setcounter{section}{1}
\setcounter{equation}{0}
\section{Notations}

It is well-known that unlike a purely bosonic algebra 
a superalgebra admits different Weyl inequivalent
choices of simple root systems, which correspond to inequivalent Dynkin 
diagrams. In the case of $osp(2|2)$, one has two choices of simple roots
which are unrelated by Weyl transformations: a system of fermionic and
bosonic simple roots (i.e. the so-called standard basis), or a purely
fermionic system of simple roots (that is the so-called non-standard
basis). So it is useful to obtain results in the two different bases
for different physical applications. Moreover, it seems that
only in the non-standard basis  could $osp(2|2)$ be twisted to give 
$osp(2|2)^{(2)}$. 

\subsection{$osp(2|2)^{(1)}$ in the standard basis}
 
Let $E\;(F)$ and $e\;(f)$ be the generators 
corresponding to the even and odd simple roots of $osp(2|2)$
in the standard (distinguished) basis, 
respectively. Let $\bar{e},\; \bar{f}$ be the odd non-simple generators. 
They satisfy the following (anti-)commutation relations: 
\beqa
 &&[E,F]=H, ~~~~[H,E]=2E,~~~~[H,F]=-2F, \n
 &&\{e,f\}=-\frac{1}{2}(H-H^{\prime}), ~~~~[H,e]=-e,~~~~[H,f]=f, \n
 &&[H^{\prime},e]=-e,~~~~[H^{\prime},f]=f,\n
 &&[E,e]=\bar{e},~~~~[F,f]=\bar{f}, \n
 &&\{\bar{e},\bar{f}\}=-\frac{1}{2}(H+H^{\prime}),\n
 &&\{e,\bar{f}\}=-F,~~~~\{\bar{e},f\}=E, \n
 &&[E,\bar{f}]=f,~~~~[F,\bar{e}]=e,\n
 &&[H,\bar{e}]=\bar{e},~~~~[H,\bar{f}]=-\bar{f},\n
 &&[H^{\prime},\bar{e}]=-\bar{e},~~~~[H^{\prime},\bar{f}]=\bar{f}.
\eeqa

\no All other (anti-)commutators are zero. The quadratic Casimir 
is given by
\beq
C_2=\frac{1}{2}\left(H(H+2)- H^{\prime}(H^{\prime}+2)\right) 
+ 2fe -2 {\bar f}{\bar e}+2FE.
\eeq

\no This quadratic Casimir is useful in sequel to 
construct the energy-momentum tensor. 

The current superalgebra $osp(2|2)^{(1)}_k$ in the standard basis
can be written as 
\beq
J_A (z) J_B (w)=k\frac{str(AB)}{(z-w)^2} +
 f_{AB}^{~~~C}\frac{J_C(w)}{(z-w)},\label{current-s}
\eeq
where $f_{AB}^{~~~C}$ are structure constants related to generators 
$A,B$ and $C$, which can be read off from the above (anti-)commutation
relations.

\subsection{$osp(2|2)^{(1)}$ in the non-standard basis}

In the non-standard basis, simple roots of $osp(2|2)$ are all fermionic.
Let $e,\;f,\;\bar{e},\;\bar{f}$ be the generators corresponding such
fermionic simple roots, and let $E,\;F$ be the non-simple generators.
They obey the (anti-) commutation relations:
\beqa
 &&\{e,f\}=-\frac{1}{2}(H-H^{\prime}), ~~~~[H,e]=e,~~~~[H,f]=-f, \n
 &&[H^{\prime},e]=e,~~~~[H^{\prime},f]=-f,\n
 &&[H,\bar{e}]=\bar{e},~~~~[H,\bar{f}]=-\bar{f},\n
 &&[H^{\prime},\bar{e}]=-\bar{e},~~~~
    [H^{\prime},\bar{f}]=\bar{f},\n
 &&\{\bar{e},\bar{f}\}=-\frac{1}{2}(H+H^{\prime}),\n
 &&\{e,\bar{e}\}=E,~~~~\{\bar{f},f\}=-F, \n
 &&[E,F]=H, ~~~~[H,E]=2E,~~~~[H,F]=-2F, \n
 &&[E,f]=\bar{e},~~~~[F,e]=\bar{f}, \n
 &&[E,\bar{f}]=e,~~~~[F,\bar{e}]=f.  \label{cr-nst}
\eeqa

\no All other (anti-)commutators are zero, and the 
quadratic Casimir is 
\beq
C_2=\frac{1}{2}\left(H^2- H^{\prime~2}\right) 
-2fe -2 {\bar f}{\bar e}+2FE.
\eeq

\no The current superalgebra $osp(2|2)^{(1)}_k$ in the non-standard
basis has the similar form as (\ref{current-s}) except that 
$f_{AB}^{~~~C}$ are now derived from (\ref{cr-nst}).

\subsection{Twisted superalgebra $osp(2|2)^{(2)}$ }

Let us start with some basics of twisted affine algebras~\cite{Kac}. 
Let $g$ be a simple finite-dimensional Lie algebra and $\sgm $ be an 
automorphism of $g$ satisfying $\sgm ^r =1$ for a positive integer $r$, 
then $g$ can be decomposed into the form: 
$g=\oplus_{j=0}^{r-1}~ g_j$, 
where $g_j$ is the eigenspace of $\sgm $ with eigenvalue 
$e^{2j{\pi} i/r}$, and $[g_i, g_j]\subset g_{(i+j)~mod~r}$~, 
then $r$ is called the order of the automorphism. 

Here we only consider the simplest twisted affine 
superalgebra $osp(2|2)^{(2)}$ so that $g=osp(2|2)$ and $r=2$. 
We can write  
\beq
osp(2|2)=g_0\op g_1
\eeq

\no where $g_0=osp(1|2)$ is a fixed point sub-superalgebra under the 
automorphism, while $g_1$ is a three dimensional representation of $g_0$, $g_0$ 
and $g_1$ satisfy $[g_i, g_j]\subset g_{(i+j)~mod ~2}$. We denote
the basis of $g_0$ by $e,\;f,\;E,\;F,\;H$
and the basis for $g_1$ by $e',\;f,\;H'$.
The commutation relations of $osp(2|2)$ in this basis are
\beqa
 &&[E,F]=H, ~~~[H,E]=2E,~~~[H,F]=-2F, \n
 && \{e,e\}=2E,~~~\{f,f\}=-2F,~~~\{e,f\}=H, \n
 &&[E,f]=-e,~~~[F,e]=-f, \n
 &&[H,e]=e,~~~[H,f]=-f, \n
 && \{ e^{\prime},e^{\prime} \}=-2E,~~~
    \{ f^{\prime},f^{\prime} \}=2F,~~~
    \{ e^{\prime},f^{\prime}\}=H,  \n
 && [H^{\prime},e^{\prime}]=-e,~~~
    [H^{\prime},f^{\prime}]=-f,\n
 && \{ e^{\prime},f \}=H^{\prime},~~~
    \{ f^{\prime},e \}=H^{\prime}, \n
 && [H,e^{\prime}]=e^{\prime},~~~
    [H,f^{\prime}]=-f^{\prime},\n
 && [E,f^{\prime}]=e^{\prime},~~~
    [F,e^{\prime}]=f^{\prime},\n
 && [H^{\prime},e]=-e^{\prime},~~~
    [H^{\prime},f]=f^{\prime}.\label{cr-twisted}
\eeqa

\no All other (anti-)commutators are zero, and the quadratic Casimir is  
\beq
C_2=\frac{1}{2}\left(H^2- H^{\prime~2}\right) 
+2fe +2f^{\prime} e^{\prime} +2FE.
\eeq

\no The current superalgebra $osp(2|2)^{(2)}_k$ reads 
\beq
J_A (z) J_B (w)=k\frac{str(AB)}{(z-w)^2} +
 f_{AB}^{~~~C}\frac{J_C(w)}{z-w},\label{current-twisted}
\eeq
where $f_{AB}^{~~~C}$ are read off from (\ref{cr-twisted}).

\setcounter{section}{2}
\setcounter{equation}{0}
\section{Wakimoto realizations and screen currents}

In this section we examine the free field realizations of
$osp(2|2)^{(1)}_k$ and $osp(2|2)^{(2)}_k$ and their screen
currents. The results for $osp(2|2)^{(2)}_k$ are new.
For $osp(2|2)^{(1)}_k$ similar results have also been obtained
in \cite{BOo,Ito,Bow96,Ras98} using different approaches. Let us
remark that free field realizations of affine $osp(2|2)$ at
$k=1$ have also been constructed in \cite{Sha98,Lud00}. (Note:
Ludwig used a different convention for $osp(2|2)^{(1)}_k$ in
\cite{Lud00}. $k=-2$ in his convention is equivalent to $k=1$ in
our convention.) 

To obtain free field realizations 
we first construct Fock space representations of $osp(2|2)$ 
corresponding to the bases given in section $2$. 
Let $E_\alpha$ denote the raising generators of $osp(2|2)$. 
A highest weight state $|P,Q,P>$ of $osp(2|2)$ is defined by  
\beq
E_\alpha|P,Q,P>=0,~~~~ 
H|P,Q,P>=P|P,Q,P>,~~~~H^{\prime}|P,Q,P>=Q|P,Q,P>.
\eeq

\subsection{$osp(2|2)^{(1)}_k$ in the standard basis}

Let $A_{st}=xF+{\tht} f+\bar{\tht}\bar{f}$ be an operator in the
standard basis of $osp(2|2)$,
where $x$ is bosonic coordinate, ${\tht}$ and $\bar{\tht}$ are 
fermionic coordinates. The action of $e^{A_{st}}$ on the highest weight 
state $|p,q,p>$ generates a coherent state of $osp(2|2)$. Write
\beqa
&&b_ge^{A_{st}}|p,q,p>=D_ge^{A_{st}}|p,q,p>,\n
&&f_ge^{A_{st}}|p,q,p>=d_ge^{A_{st}}|p,q,p>,
\eeqa 

\no for bosonic generators $b_g$ and fermionic generators $f_g$ of
$osp(2|2)$. Here $D_g$, $d_g$ are the corresponding differential operators. 
Using the Baker-Campbell-Hausdorff (BCH) formula and $osp(2|2)$
commutation relations, we obtain
\beqa
&&d_{\bar{f}}=\p _{\bar{\tht}}, \n
&&d_f=\p _{\tht}+\frac{1}{2}x\p _{\bar{\tht}}, \n
&&D_F=\p _{x}-\frac{1}{2}\tht \p _{\bar{\tht}}, \n
&&D_H=p+2x\p _{x}- \tht \p _{\tht}+{\bar{\tht}}\p _{\bar{\tht}},\n
&&D_{H^{\prime}}=q- \tht \p _{\tht}-{\bar{\tht}}\p _{\bar{\tht}},\n
&&d_{e}=-\frac{1}{2}(p-q)\tht - {\bar\tht} \p _{x}
-\frac{1}{2}{\tht} x\p _{x}
-\frac{1}{2}{\tht \bar{\tht}} \p _{\bar{\tht}},\n
&&D_{E}=-px - {\bar{\tht}} \p _{\tht} -x^2\p _{x}
-\frac{1}{2}x{\bar{\tht}} \p _{\bar{\tht}}
+\frac{1}{2}x{\tht} \p _{\tht}
-\frac{1}{4}x^2 {\tht} \p _{\bar{\tht}},\n
&&d_{\bar{e}}=-\frac{1}{2}(p+q){\bar{\tht}}
-\frac{1}{4}(3p-q)x{\tht} 
-{\bar{\tht}}x \p _{x} 
+{\bar{\tht}} \tht \p _{{\tht}}
-\frac{1}{2}{\tht} x^2 \p _{x}.\label{diff-st}
\eeqa  

\no It is straightforward to prove that the above differential 
operators satisfy the algebraic relations of $osp(2|2)$ in the standard
basis. 

We now use the differential operator representation
(\ref{diff-st}) to find 
the Wakimoto realization of $osp(2|2)^{(1)}_k$ in terms of 
one bosonic $\bt$-$\gm$ pair, two fermionic $b$-$c$ type systems
and two free scalar fields. These free fields have the following OPEs: 
\beqa
&&\bt (z)\gm (w)=-\gm(z)\bt(w)=-\frac{1}{z-w}, ~~~
\ps (z) \psd (w)=\psd(z)\ps(w)=-\frac{1}{z-w},\n
&&\psb (z) \psbd (w)=\psbd(z)\psb(w)=-\frac{1}{z-w},~~~~
\ph (z) \ph (w)=-ln (z-w)= \ph ^{\prime} (z)\ph ^{\prime} (w).\n
\eeqa

The free field realization of $osp(2|2)^{(1)}$ in the standard basis is
obtained by the following substitution:
\beqa
&&d_{f(\bar{f})}\rightarrow j_{e(\bar{e})}(z),~~~~
d_{e(\bar{e})}\rightarrow j_{f(\bar{f})}(z),~~~~
D_F\rightarrow J_{E}(z),\n
&&D_E\rightarrow J_F(z),~~~~D_{H(H')}\rightarrow
J_{H(H')}(z),\n
&&\partial_x\rightarrow \beta(z),~~~~x\rightarrow\gamma(z),~~~~
\theta\;(\bar{\theta})\rightarrow \psi(z)\;(\bar{\psi}(z)),\n
&&\partial_{\theta\,(\bar{\theta})}\rightarrow
\psi^\dagger(z)\;(\bar{\psi}^\dagger(z)),~~~~p\rightarrow
i\alpha_+\partial\phi(z),~~~~q\rightarrow
i\alpha_+\partial\phi'(z)\label{substitution}
\eeqa
in the differential operator realization (\ref{diff-st}) and a
subsequent addition of anomalous terms linear in $\partial
\psi(z),~ \partial\gamma(z)$ or $\partial\bar{\psi}(z)$ in
currents $j_{f\,(\bar{f})}(z)$ and $J_F(z)$. The result is
\beqa
&&j_{\bar{e}}(z)=\psbd (z), \n
&&j_e(z)=-\psd (z)-\frac{1}{2}\gm (z) \psbd (z), \n
&&J_E(z)=\bt (z)-\frac{1}{2}\ps (z) \psbd (z), \n
&&J_H(z)=i\al _{+}\p \ph (z)+2\bt (z)\gm (z)- \ps (z) \psd (z) 
+\psb (z) \psbd (z),\n
&&J_{H^{\prime}}(z)=\al _{+} \p \ph ^{\prime} (z)
- \ps (z) \psd (z) - \psb(z) \psbd (z),\n
&&j_{f}(z)=-\frac{1}{2}\al _{+} (i\p \ph (z)-\p \ph ^{\prime} (z))\ps (z)
 -\bt (z) \psb (z) -\frac{1}{2}\bt (z)\gm (z) \ps (z) \n
&&~~~~~~~~~-\frac{1}{2} \psb(z) \psbd (z) \ps (z) +(k+\frac{1}{2})\p \ps (z),\n
&&J_{F}(z)=-i\al _{+}\p \ph (z) \gm (z) -\bt (z)\gm ^2 (z)
-\psb (z) \psd (z) 
+\frac{1}{2}\gm(z) (\ps (z) \psd (z) + \psb(z) \psbd (z) ) \n
&&~~~~~~~~~~-\frac{1}{4}\gm ^2 (z)\ps (z) \psbd (z) 
-(k-\frac{1}{2}) \p \gm (z),\n
&&j_{\bar{f}}(z)=\frac{1}{2}\al _{+}(i\p \ph (z)
+\p \ph ^{\prime} (z) )\psb (z)
+\frac{1}{4}\al _{+}(3i\p \ph (z)-\p \ph ^{\prime} (z) )\gm (z) \ps (z) \n 
&&~~~~~~~~~+\bt (z)\gm (z)\psb (z) - \ps (z) \psd (z) \psb(z)
+\frac{1}{2}\bt (z)\gm ^2 (z) \ps (z) \n
&&~~~~~~~~~+k \p \psb (z) +\frac{1}{2}(k-1)\ps (z) \p \gm (z)
-\frac{1}{2}(k+1)\gm (z) \p \ps (z),\label{free-st}
\eeqa  

\no where $\al _+ =\sqrt{2k+2}$, and normal ordering is implied 
in the expressions.  

The energy-momentum tensor is obtained by the Sugawara construction. 
Due to singularities which arise when multiplying currents at the same
point we need to consider a regularization to remove such divergences.
We use the usual point-splitting regularization where singular parts
appearing in the OPEs of the currents are subtracted. This is equivalent
to defining the normal ordered product of two fields $A(z)$ and $B(z)$
by
\beq
:AB:(z)\equiv \oint_w \frac{dz}{2\pi i}\frac{A(z)B(w)}{z-w}.
\eeq
In the present case, the Sugawara energy-momentum tensor is given by
\beqa
T_{st}(z)&=&\frac{1}{2(k+1)}:\left(
       \frac{1}{2}J_H(z)J_H(z)-\frac{1}{2}J_{H'}(z)J_{H'}(z)
       +J_E(z)J_F(z)+J_F(z)J_E(z)\right.\n
& &\left.-j_e(z)j_f(z)+j_f(z)j_e(z)+j_{\bar{e}}(z)j_{\bar{f}}(z)
   -j_{\bar{f}}(z)j_{\bar{e}}(z)\right):.
\eeqa
By means of the free field representation of the currents, we get
\beqa
&&T_{st}(z)=-\bt (z) \p \gm (z) +\psd (z) \p \ps (z) +\psbd(z) \p \psb (z) \n
&&~~~~~~~~~+\frac{1}{2}\left ( (i\p \ph (z))^2 
+(\p \ph ^{\prime} (z))^2 \right)
-\frac{1}{\al _+}\left ( i\p ^2\ph (z) 
-\p ^2\ph ^{\prime} (z)\right).
\eeqa

\no The energy-momentum tensor satisfies the OPE, 
\beq
T_{st}(z)T_{st}(w)=\frac{2T_{st}(w)}{(z-w)^2}+\frac{\p T_{st}(w)}{z-w)}.
\eeq

\no So the Virasoro central charge of the theory is zero.

An important object in the free field approach is screening current. 
Screening currents are  primary fields with conformal dimension $1$, 
and their integrations give the screening charges. They communite with the 
affine currents up to a total derivative. These properties ensure that 
screening charges may be inserted into correlators while the conformal or 
affine ward identities remain intact. For the present case,
the screening currents of first kind are found to be
\beqa
&&s_{s,1}(z)=\left(\psd (z)-\frac{1}{2}\gm (z) \psbd (z)\right)
{\rm exp}\{\frac{1}{\al _+}\left ( i \ph (z) 
-\ph ^{\prime} (z)\right)\}, \n
&&s_{s,2} (z)=\left( \bt (z)+ \frac{1}{2}\ps (z) \psbd (z)\right) 
{\rm exp}\{-\frac{2}{\al _+} i \ph (z)\} \label{screen1-st} 
\eeqa
and the screening current of second kind is 
\beqa
&&s_{II} (z)=\left( \bt ^{-k-1}(z)
- \frac{k+1}{2}\bt ^{-k-2}(z)\ps (z) \psbd (z)\right) 
{\rm exp}\{\al _+ i \ph (z)\}. \label{screen2-st} 
\eeqa



Similar results to (\ref{free-st}, \ref{screen1-st}, \ref{screen2-st}) 
have also been obtained in \cite{Bow96,Ras98} using different approaches.


\subsection{$osp(2|2)^{(1)}_k$ in the non-standard basis}

In the non-standard basis, the action of the operator $e^{A_{nst}}$
with $A_{nst}=xF+{\tht} f+\bar{\tht}\bar{f}$ on the highest weight 
state $|p,q,p>$ generates a coherent state of $osp(2|2)$,
where $x$ is bosonic coordinate, ${\tht}$ and $\bar{\tht}$ are 
fermionic coordinates. Denote the action of $osp(2|2)$
generators on this coherent state by
\beqa
&&b_ge^{A_{nst}}|p,q,p>=D_ge^{A_{nst}}|p,q,p>,\n
&&f_ge^{A_{nst}}|p,q,p>=d_ge^{A_{nst}}|p,q,p>,
\eeqa 

\no where $b_g$, $f_g$ are the bosonic and 
fermionic generators of $osp(2|2)$, respectively, and 
$D_g$, $d_g$ are the corresponding differential operators.
After some algebraic manipulations, we find
\beqa
&&d_{\bar{f}}=\p _{\tht}-\frac{1}{2}\bar{\tht}\p _x, \n
&&d_f=\p _{\bar {\tht}}-\frac{1}{2}{\tht}\p _x, \n
&&D_F=\p _{x}, \n
&&D_H=p+2x\p _{x}+ \tht \p _{\tht}+{\bar{\tht}}\p _{\bar{\tht}},\n
&&D_{H^{\prime}}=q+\tht \p _{\tht}-{\bar{\tht}}\p _{\bar{\tht}},\n
&&d_{e}=-\frac{1}{2}(p-q)\tht +x\p _{\bar\tht}
-\frac{1}{2}{\tht} x\p _{x}
-\frac{1}{2}{\tht \bar{\tht}} \p _{\bar{\tht}},\n
&&d_{\bar{e}}=-\frac{1}{2}(p+q){\bar{\tht}}
+x \p _{\tht}
-\frac{1}{2}{\bar{\tht}} \tht \p _{{\tht}}
-\frac{1}{2}{\bar{\tht}} x \p _{x}, \n
&&D_{E}=-px +\frac{1}{2}q {\tht}{\bar{\tht}} -x^2\p _{x}
-x{\bar{\tht}} \p _{\bar{\tht}}
-x{\tht} \p _{\tht}.\label{diff-nst}
\eeqa  

\no It is easy to show that the above differential operators 
give a realization of $osp(2|2)$ in the non-standard basis. 

With the help of the differential operator representation
(\ref{diff-nst})  and by a subsitution similar to
(\ref{substitution}) and an addition of suitable anomalous
terms we find the free field realization of $osp(2|2)^{(1)}_k$ 
in the non-standard basis
\beqa
&&J_{E}(z)=\bt (z), \n
&&j_e(z)=\psd (z)-\frac{1}{2}\bt (z) \psb (z), \n
&&j_{\bar{e}}(z)=\psbd (z)-\frac{1}{2}\bt (z)\ps (z), \n
&&J_H(z)=i\al _{+}\p \ph (z)+2\bt (z)\gm (z)+ \ps (z) \psd (z) 
+\psb (z) \psbd (z),\n
&&J_{H^{\prime}}(z)=\al _{+} \p \ph ^{\prime} (z)
+ \ps (z) \psd (z) - \psb(z) \psbd (z),\n
&&j_{f}(z)=\frac{1}{2}\al _{+} (i\p \ph (z)-\p \ph ^{\prime} (z))\ps (z)
 -\gm (z) \psbd (z) +\frac{1}{2}\bt (z)\gm (z) \ps (z) \n
&&~~~~~~~~~+\frac{1}{2} \ps(z) \psb (z) \psbd (z) 
+(k+\frac{1}{2})\p \ps (z),\n
&&j_{\bar{f}}(z)= \frac{1}{2}\al _{+} 
(i\p \ph (z)+\p \ph ^{\prime} (z))\psb (z) -\gm (z) \psd (z)
+\frac{1}{2}\bt (z) \gm (z)\psb (z)  \n
&&~~~~~~~~~~+\frac{1}{2}\psb (z) \ps (z)\psd (z) ) 
+(k+\frac{1}{2}) \p \psb (z),\n
&&j_{F}(z)=-\al _{+}i\p \ph (z) \gm (z)
+\frac{1}{2}\al _{+}\p \ph ^{\prime} (z)\ps (z) \psb (z) 
-\bt (z) \gm ^2 (z) \n 
&&~~~~~~~~~~+\gm (z)\left(\ps (z)\psd (z) +\psb(z)\psbd (z) \right) 
-k \p \gm (z)  \n
&&~~~~~~~~~~+\frac{1}{2}(k+1)
\left(\psb (z) \p \ps (z) +\ps (z) \p \psb (z) \right).\label{free-nst}
\eeqa  

\no The energy-momentum tensor in the non-standard basis is given by
\beqa
&&T_{nst}(z)=-\bt (z) \p \gm (z) +\psd (z) \p \ps (z) +\psbd(z) \p \psb (z) \n
&&~~~~~~~~~+\frac{1}{2}\left ( (i\p \ph (z))^2 
-(\p \ph ^{\prime} (z))^2 \right).
\eeqa
This energy-momentum tensor has no terms with background charges,
and obeys the OPE
\beq
T_{nst}(z)T_{nst}(w)=\frac{2T_{nst}(w)}{(z-w)^2}+\frac{\partial
T_{nst}(w)}{z-w}.
\eeq
The sceening currents in the non-standard basis are
\beqa
&&s_{n,1}(z)=\left(\psd (z)+\frac{1}{2}\bt (z) \psb (z)\right)
{\rm exp}\{-\frac{1}{\al _+}\left ( i \ph (z) 
-\ph ^{\prime} (z)\right)\}\n
&&s_{n,2}(z)=\left( \psbd (z)+ \frac{1}{2}\bt (z) \ps (z)\right) 
{\rm exp}\{-\frac{1}{\al _+} \left(i \ph (z) 
+\ph ^{\prime} (z)\right)\}. \label{screen-nst} 
\eeqa

%


Similar results as (\ref{free-nst}, \ref{screen-nst}) have also been given in
\cite{Ito,Ras98}, though based on different approaches and conventions.

\subsection{$osp(2|2)^{(2)}_k$}

Here $osp(2|2)$ is decomposed into $g_0\op g_1$. 
The action of $e^{A_t}$
with $A_t=xF+{\tht} f+{\tht} ^{\prime} f^{\prime} $ on 
the highest weight state $|p,q',p>$ generates a coherent state 
of $osp(2|2)$ in the basis given by (\ref{cr-twisted}), 
where $x$ is bosonic coordinate, ${\tht}$ and 
${\tht}^{\prime}$ are fermionic coordinates. Again write
\beqa
&&b_ge^{A_t}|p,q^{\prime},p>=D_ge^{A_t}|p,q^{\prime},p>,\n
&&f_ge^{A_t}|p,q^{\prime},p>=d_ge^{A_t}|p,q^{\prime},p>,
\eeqa 
where $b_g, f_g$ stand for the bosonic and fermionic generators of
$osp(2|2)$ in the basis (\ref{cr-twisted}), and $D_g, d_g$ are the
corresponding differential operators. After some algebraic computations
we get
\beqa
&&D_F=\p _{x}, \n
&&d_f=\p _{ {\tht}}-{\tht}\p _x, \n
&&d_{{f}^{\prime}}=\p _{{\tht}^{\prime}}+ {\tht}^{\prime}\p _x, \n
&&D_H=p+2x\p _{x}+ \tht \p _{\tht}+
{{\tht}^{\prime}}\p _{{\tht}^{\prime}},\n
&&D_{H^{\prime}}=q^{\prime} -\tht \p _{{\tht}^{\prime}}
-{{\tht}^{\prime}}\p _{{\tht}},\n
&&d_{e}=p\tht + q^{\prime}{\tht}^{\prime} 
- x \p _{\tht}+{\tht} x\p _{x}
+{\tht {\tht}^{\prime}} \p _{{\tht}^{\prime}},\n
&&D_{E}=-px +q^{\prime} {\tht}{{\tht}^{\prime}} -x^2\p _{x}
-x{\tht} \p _{\tht}-x{{\tht}^{\prime}} \p _{{\tht}^{\prime}}, \n
&&d_{e^{\prime}}=p{\tht}^{\prime} + q^{\prime}{\tht} 
+ x \p _{{\tht}^{\prime}}+{{\tht}^{\prime}} x\p _{x}
+{\tht}^{\prime} \tht \p _{\tht}.\label{diff-t}
\eeqa  

\no Indeed these differential operators satisfy the algebraic 
relations (\ref{cr-twisted}). 

By means of the differential operator realization
(\ref{diff-t}) and a substitution similar to
(\ref{substitution}) and an addition of suitable anomalous
terms we find the free field realization of $osp(2|2)^{(2)}_k$
\beqa
&&J_{E}(z)=\bt (z), \n
&&j_e(z)=\psd (z)-\bt (z) \ps (z), \n
&&j_{e^{\prime}}(z)=\pspd (z)+\bt (z)\psp (z), \n
&&J_H(z)=i\al _{+}\p \ph (z)+2\bt (z)\gm (z)+ \ps (z) \psd (z) 
   +\psp (z) \pspd (z),\n
&&J_{H^{\prime}}(z)=\al _{+} \p \ph ^{\prime} (z)
- \ps (z) \pspd  (z) - \psp(z) \psd (z),\n
&&j_{f}(z)=-\al _{+} \left(i\p \ph (z) \ps (z)
+\p \ph ^{\prime} (z)\psp (z)\right)
 +\gm (z) \psd (z) -\bt (z)\gm (z) \ps (z) \n
&&~~~~~~~~~- \ps(z) \psp (z) \pspd (z) 
-(2k+1)\p \ps (z),\n
&&j_{f^{\prime}}(z)=-\al _{+} \left(i\p \ph (z) \psp (z)
  +\p \ph ^{\prime} (z)\ps (z)\right)
 -\gm (z) \pspd (z) -\bt (z)\gm (z) \psp (z) \n
&&~~~~~~~~~- \psp(z) \ps (z) \psd (z) 
-(2k+1)\p \psp (z),\n
&&J_{F}(z)=-\al _{+} \left(i\p \ph (z) \gm (z)
-\p \ph ^{\prime} (z) \ps (z) \psp (z)\right)
 -\bt (z)\gm ^2(z) \n 
&&~~~~~~~~~-\gm(z)\left( \ps (z)\psd (z) 
+ \psp (z) \pspd (z)\right) -k \p \gm (z)  \n
&&~~~~~~~~~~+(k+1)
\left(\ps (z) \p \ps (z) -\psp (z) \p \psp (z) \right),
\eeqa  
where $\psi'(z)$ and $\psi^{'\dagger}(z)$ are free fermionic
fields having the OPEs
\beq
\psi'(z)\psi^{'\dagger}(w)=\psi^{'\dagger}(z)\psi'(w)=-\frac{1}{z-w}.
\eeq
It is straightforward to check that the above currents satisfy
the OPEs of $osp(2|2)^{(2)}_k$ given in last section.

The energy-momentum tensor is 
\beqa
&&T_t(z)=-\bt (z) \p \gm (z) +\psd (z) \p \ps (z) 
       +\pspd(z) \p \psp (z) \n
&&~~~~~~~~~+\frac{1}{2}\left ( (i\p \ph (z))^2 
-(\p \ph ^{\prime} (z))^2 \right).
\eeqa
There are no background charges in the expression of the energy-momentum
tensor, and its OPE reads
\beq
T_t(z)T_t(w)=\frac{2T_t(w)}{(z-w)^2}+\frac{\partial T_t(w)}{z-w}.
\eeq
So we are dealing with a conformal field theory with zero Virasoro
central charge.


It is first pointed in~\cite{Din01} that in twisted case, the usual 
method to derive the screening currents is inappropriate. The screening 
currents should be twisted. The twisted screening currents for 
$osp(2|2)^{(2)}$ are found to be
\beqa
&&s_{t,+}(z)=\left(\psd (z)+\pspd (z)
+ \bt (z) \ps (z)-\bt (z) \psp (z)\right)
{\rm exp}\{-\frac{1}{\al _+}\left ( i \ph (z) 
+\ph ^{\prime} (z)\right)\} \n
&&~~~~~~~=\left(\psd (z)+\pspd (z)
+ \bt (z) \ps (z)-\bt (z) \psp (z)\right)
{\tl s_{t,+}}(z), \n
&&s_{t,-} (z)=\left(\psd (z)-\pspd (z)
+ \bt (z) \ps (z)+\bt (z) \psp (z)\right)
{\rm exp}\{-\frac{1}{\al _+}\left ( i \ph (z) 
-\ph ^{\prime} (z)\right)\} \n
&&~~~~~~~=\left(\psd (z)-\pspd (z)
+ \bt (z) \ps (z)+\bt (z) \psp (z)\right)
{\tl s_{t,-}}(z).
\eeqa
These screening currents satisfy the OPEs:
\beq
j_e (z) s_{t,\pm} (w)=J_E (z) s_{t,\pm} (w)
=j_{\bar {e}} (z) s_{t,\pm} (w)=J_H (z) s_{t,\pm}(w)
=J_{H^{\prime}}(z) s_{t,\pm}(w)=\cdots,
\eeq

\no and 
\beqa
&&j_f (z) s_{t,\pm}(w)=
-\p _w \left( \frac{\al ^2 _{+}}{z-w}
{\tl s_{t,\pm}} (w)\right), \n
&&j_{f^{\prime}} s_{t,\pm}(w)=
-\p _w \left( \frac{\al ^2 _{+}}{z-w}
{\tl s_{t,\pm}} (w)\right), \n
&&J_F (z) s_{t,+}(w)=\p _w \left(
\frac{\al _{+}^2}{z-w}
( \ps (w)-\psp (w) ) {\tl s}_{t,+}(w) \right), \n
&&J_F (z) s_{t,-}(w)=\p _w \left( 
\frac{\al _{+}^2}{z-w}
( \ps (w)+\psp (w) ){\tl s}_{t,-}(w) \right).
\eeqa

\no There does not seem to have screening current of the second kind for
$osp(2|2)^{(2)}_k$.

\setcounter{equation}{0}
\section{Primary fields}

Primary fields are fundamental objects in conformal field theories. 
A primary field $\Ps$ has the following OPE with the energy-momentum 
tensor:    
\beq
T(z) \Ps (w)=\frac{\Delta_{\Ps}}{(z-w)^2}\Ps (w)
+\frac{\p _w \Ps (w)}{z-w}+\ldots,
\eeq

\no where the $\Delta_{\Ps}$ is the conformal dimension of $\Ps$. Moreover 
the OPEs of $\Ps$ with the affine currents do not contain poles higher
than first order. A special kind of the primary fields is highest 
weight state. 

Let us remark that certain representations were investigated 
for $osp(2|2)$ in \cite{Dob93,Kob94} and for $osp(2|2)^{(1)}$ in
\cite{Bow97,Sem97}. Here we are concerned with primary fields, which
requires the construction of all representations.

\subsection {$osp(2|2)^{(1)}$ primary fields in the standard basis}

It is easy to see that the highest weight state of the algebra is 
\beq
V_{p,q}(z)={\rm exp} \{\frac{2}{\al _+}(p i \ph (z)- q \ph ^{\prime}(z))\}
\eeq

\no where $p,~q$  are given complex numbers labelling the
representation. The conformal dimension of the field is 
\beq
\Delta_{p,q}=\frac{p(p+1)-q(q+1)}{k+1}.
\eeq
If $q\neq p, -p-1$, then $\Delta_{p,q}\neq 0$ and the corresponding
representations are typical. When $q=p, -p-1$, we have $\Delta_{p,q}=0$
and atypical representations arise.
In order for the representation to be finite-dimensional,
we find that $p$ must be an integer or half-integer. 
The full bases of the representation labelled by $p,q$ are
\beqa
S^m _{p,q}(z)&=&\left(-\gm (z)\right)^{p-m}V_{p,q}(z), 
~~m=p,~p-1,\cdots, -(p-1),-p, \n
s^n _{p,q}(z)&=&\left(-\gm (z)\right)^{(p-3/2)-n}
\left(\psb (z)+\frac{1}{2}\gm (z) \ps (z) \right)V_{p,q}(z),\n 
& &n=(p-3/2),\cdots,-(p+1/2),~~p\geq 1/2, \n
{\tl s}^l _{p,q}(z)&=&(p-q)\left(-\gm (z)\right)^{(p-3/2)-l}
\left(\psb (z)-\frac{1}{2}\gm (z) \ps (z) \right)V_{p,q}(z), \n
& &l=(p-3/2),\cdots,-(p+3/2), ~~p\geq 3/2,\n
\ph _{p,q}(z)&=&(p-q)\ps (z)V_{p,q}(z), \n
{\cal S}^s _{p,q}(z)&=&(p-q)\left(-\gm (z)\right)^{(p-2)-s}
\ps (z) \psb (z) V_{p,q}(z),\n 
& &s=(p-2),\cdots,-(p+2),~~p\geq 2. 
\eeqa 

The dimensions of $S^m _{p,q}(z)$ and 
$s^n _{p,q}(z)$ are $2p+1$ and $2p$, respectively. On the other hand 
both ${\tl s}^{l} _{p,q}(z)$ and ${\cal S}^{s} _{p,q}(z)$ has $(2p+1)$ 
independent components. Note that $\ph _{p,q} (z)$ is 
one-dimensional. So the dimension of a typical represention 
(where $q\not= p, -p-1$) is $8p+4$. For an atypical representation
corresponding to $q= p$, 
$S^m _{p,q}(z)$ and $s^n _{p,q}(z)$ are the only non-vanishing fields
and so the dimension of the atypical representation is $4p+1$. 

By means of the free field representations given in section 3, we
compute the OPEs of affine currents with the primary fields.
The OPEs of the $osp(2|2)$ currents with $S^m _{p,q}(z)$ are 
\beqa
&&J_E (z) S^m _{p,q}(w)= \frac{p-m}{z-w}S^{m+1} _{p,q}(w), \n
&&J_F (z) S^m _{p,q}(w)= \frac{p+m}{z-w}S^{m-1} _{p,q}(w), \n
&&J_H (z) S^m _{p,q}(w)= \frac{2m}{z-w}S^{m} _{p,q}(w), \n
&&J_{H^{\prime}} (z) S^m _{p,q}(w)= \frac{2q}{z-w}S^{m} _{p,q}(w), \n
&&j_e (z) S^m _{p,q}(w)=0, \n
&&j_{\bar e} (z) S^m _{p,q}(w)=0, \n
&&j_f (z) S^m _{p,q}(w)= \frac{1}{z-w}
\left( (m-q)s^{m-1/2} _{p,q}(w)-{\tl s}^{m-1/2} _{p,q}(w)\right),~~ 
m\leq (p-1), \n
&&j_f (z) S^p _{p,q}(w)= \frac{-1}{z-w}\ph _{p,q} (w), \n
&&j_{\bar f }(z) S^m _{p,q}(w)= \frac{1}{z-w}
\left( (p+m)s^{m-3/2} _{p,q}(w)-{\tl s}^{m-3/2} _{p,q}(w)\right).
\eeqa

\no When $q=p$, terms involving $\phi_{p,q}(z)$ and
${\tl s}^l _{p,q}(z)$ disappear. The OPEs with  $s^n _{p,q}(z)$ are 
\beqa
&&J_E (z) s^n _{p,q}(w)= \frac{(p-3/2)-n}{z-w}s^{n+1} _{p,q}(w), \n
&&J_F (z) s^n _{p,q}(w)= \frac{(p+1/2)+n}{z-w}s^{n-1} _{p,q}(w), \n
&&J_H (z) s^n _{p,q}(w)= \frac{2n+2}{z-w}s^{n} _{p,q}(w), \n
&&J_{H^{\prime}} (z) s^n _{p,q}(w)= \frac{1+2q}{z-w}s^{n} _{p,q}(w), \n
&&j_e (z) s^n _{p,q}(w)=-\frac{1}{z-w}S^{n+1/2} _{p,q}(w), \n
&&j_{\bar e} (z) s^n _{p,q}(w)=-\frac{1}{z-w}S^{n+3/2} _{p,q}(w), \n
&&j_f (z) s^n _{p,q}(w)=- \frac{1}{z-w}{\cal S}^{n-1/2} _{p,q}(w), \n
&&j_{\bar f }(z)s^n _{p,q}(w)= -\frac{1}{z-w}
{\cal S}^{n-3/2} _{p,q}(w).
\eeqa

\no When $q=p$, terms containing
${\cal S}^{s} _{p,q}(w)$ disappear. The relations of the currents 
with ${\tl s}^l _{p,q}(z)$ and ${\cal S}^{s} _{p,q}(w)$ are   
\beqa
&&J_E (z) {\tl s}^l _{p,q}(w)= \frac{1}{z-w}
\left( ((p-1/2)-l){\tl s}^{l+1} _{p,q}(w)-(p-q)s^{l+1} _{p,q}(w)\right),~~ 
l\leq (p-5/2),\n
&&J_E (z) {\tl s}^{p-3/2} _{p,q}(w)= \frac{1}{z-w}\ph _{p,q} (w), \n
&&J_F (z) {\tl s}^l _{p,q}(w)= 
\frac{(p+3/2)+l}{z-w}{\tl s}^{l-1} _{p,q}(w),\n
&&J_H (z) {\tl s}^l _{p,q}(w)= 
\frac{2l+2}{z-w}{\tl s}^{l} _{p,q}(w), \n
&&J_{H^{\prime}} (z) {\tl s}^l _{p,q}(w)= 
\frac{1+2q}{z-w}{\tl s}^{l} _{p,q}(w), \n
&&j_e (z) {\tl s}^l _{p,q}(w)=0, \n
&&j_{\bar e} (z) {\tl s}^l _{p,q}(w)
=-\frac{p-q}{z-w}S^{l+3/2} _{p,q}(w),\n
&&j_f (z) {\tl s}^l _{p,q}(w)= \frac{(q-1/2)-l}{z-w}
{\cal S}^{l-1/2} _{p,q}(w), \n
&&j_{\bar f }(z) {\tl s}^l _{p,q}(w)=- \frac{(p+3/2)+l}{z-w}
{\cal S}^{l-3/2} _{p,q}(w),
\eeqa

\no and 
\beqa
&&J_E (z) {\cal S}^s _{p,q}(w)
= \frac{(p-2)-s}{z-w}{\cal S}^{s+1} _{p,q}(w), \n
&&J_F (z) {\cal S}^s _{p,q}(w)
= \frac{(p+2)+s}{z-w}{\cal S}^{s-1} _{p,q}(w), \n
&&J_H (z) {\cal S}^s _{p,q}(w)
= \frac{2s+4}{z-w}{\cal S}^{s} _{p,q}(w), \n
&&J_{H^{\prime}} (z) {\cal S}^s _{p,q}(w)
= \frac{2+2q}{z-w}{\cal S}^{s} _{p,q}(w), \n
&&j_e (z) {\cal S}^s _{p,q}(w)
=\frac{1}{z-w}{\tl s}^{s+1/2} _{p,q}(w), \n
&&j_{\bar e} (z) {\cal S}^s _{p,q}(w)
=\frac{1}{z-w}\left( {\tl s}^{s+3/2} _{p,q}(w)
-(p-q)s^{s+3/2} _{p,q}(w)\right),~~s\leq (p-3),\n
&&j_{\bar e} (z) {\cal S}^{p-2} _{p,q}(w)
=\frac{1}{z-w}\ph _{p,q} (w), \n
&&j_f (z) {\cal S}^s _{p,q}(w)=0, \n 
&&j_{\bar f }(z) {\cal S}^s _{p,q}(w)= 0.
\eeqa

\no Finally OPEs involving $\ph _{p,q}(w)$ read
\beqa
&&J_E (z) \ph _{p,q}(w)= \ldots,\n
&&J_F (z) \ph _{p,q}(w)=\frac{1}{z-w}
\left( (2p+1){\tl s}^{p-3/2} _{p,q}(w)
-2p(p-q) s^{p-3/2} _{p,q}(w)\right), , \n
&&J_H (z) \ph _{p,q}(w)
= \frac{2p+1}{z-w}\ph _{p,q}(w), \n
&&J_{H^{\prime}} (z) \ph _{p,q}(w)
= \frac{1+2q}{z-w}\ph _{p,q}(w), \n
&&j_e (z) \ph _{p,q}(w)=\frac{p-q}{z-w}S^{p} _{p,q}(w), \n
&&j_{\bar e} (z) \ph _{p,q}(w)=0,\n
&&j_f (z) \ph _{p,q}(w)=0, \n 
&&j_{\bar f }(z) \ph _{p,q}(w)
= -\frac{(p+q)+1}{z-w}{\cal S}^{p-2} _{p,q}(w).
\eeqa

\subsection{$osp(2|2)^{(1)}$ primary fields in the non-standard basis}

The highest weight state of the algebra is 
\beq
V_{J,q}(z)={\rm exp} \{\frac{2}{\al _+}(J i \ph (z)
   - q \ph ^{\prime}(z))\},\label{state-nst}
\eeq

\no where $J,~q$  are given complex numbers specifying the
representation. The conformal dimension of the field is 
\beq
\Delta_{J,q}=\frac{J^2-q^2}{k+1}.
\eeq
If $q\neq\pm J$, then $\Delta_{J,q}\neq 0$ and the corresponding
representations are typical. When $q=\pm J$, atypical representations
arise. For the representation to be finite-dimensional, it turns out
that $J$ must be an integer or half-integer and moreover if $J=0$ then $q$
must also be zero.  
For $J=0=q$, the atypical representation is obviously one-dimensional. 
For $J\neq 0$, a representation labelled by $J,q$ has the
following bases:
\beqa
&&N^m _{J,q}(z)=\left[ 2J \left(-\gm (z)\right)^{J-m} 
-q(J-m) \left(-\gm (z)\right)^{J-m-1}\ps (z)\psb (z) \right]V_{J,q}(z), \n 
&&~~~~~~~~~~~~~m=J,~J-1,\cdots, -(J-1),-J, ~~~J\geq 1/2,\n
&&n^l _{J,q}(z)=(J-q)\left(-\gm (z)\right)^{(J-1/2)-l}
\ps (z)V_{J,q}(z),\n 
&&~~~~~~~~~~~~~l=(J-1/2),\cdots,-(J-1/2), ~~J\geq 1/2,\n
&&{\bar n}^l _{J,q}(z)=(J+q)\left(-\gm (z)\right)^{(J-1/2)-l}
\psb (z)V_{J,q}(z),\n 
&&~~~~~~~~~~~~~l=(J-1/2),\cdots,-(J-1/2), ~~J\geq 1/2,\n
&&{\cal N}^n _{J,q}(z)=(J^2 -q^2)\left(-\gm (z)\right)^{(J-1)-l}
\ps (z) \psb (z) V_{J,q}(z),\n 
&&~~~~~~~~~~~~~n=(J-1),\cdots,-(J-1),~~J\geq 1 .
\eeqa 

It is easy to see that  $N^m _{J,q}(z)$  and ${\cal N}^{n}
_{J,q}(z)$ have $(2J+1)$ and $(2J-1)$ independent components,
respectively, and the dimensions of
$n^l _{J,q}(z)$ and ${\bar n}^l _{J,q}(z)$ are both $2J$. 
So the dimension of a typical represention (where $q\not= \pm J$) is $8J$. 
For an atypical representation, either only $N^m _{J,q}(z)$ 
and $n^n _{J,q}(z)$ survive (when $q=-J$) or only $N^m _{J,q}(z)$ and 
${\bar n}^n _{J,q}(z)$ remain (when $q=J$). So  the dimension of the 
atypical representation is $4J+1$.

The OPEs of the $osp(2|2)$ currents with $N^m _{J,q}(z)$ are
\beqa
&&J_E (z) N^m _{J,q}(w)= \frac{J-m}{z-w}N^{m+1} _{J,q}(w), \n
&&J_F (z) N^m _{J,q}(w)= \frac{J+m}{z-w}N^{m-1} _{J,q}(w), \n
&&J_H (z) N^m _{J,q}(w)= \frac{2m}{z-w}N^{m} _{J,q}(w), \n
&&J_{H^{\prime}} (z) N^m _{J,q}(w)= \frac{2q}{z-w}N^{m} _{J,q}(w), \n
&&j_e (z) N^m _{J,q}(w)=-\frac{J-m}{z-w}
{\bar n}^{m+1/2} _{J,q}(w), \n
&&j_{\bar e} (z) N^m _{J,q}(w)=-\frac{J-m}{z-w}
n^{m+1/2} _{J,q}(w), \n
&&j_f (z) N^m _{J,q}(w)=\frac{J+m}{z-w}
{n}^{m-1/2} _{J,q}(w), \n
&&j_{\bar f} (z) N^m _{J,q}(w)=\frac{J+m}{z-w}
{\bar n}^{m-1/2} _{J,q}(w).
\eeqa

\no We see that ${n}^{l} _{J,q}(z)$ and 
${\bar n}^{l} _{J,q}(z)$ are generated from $N^m _{J,q}(z)$ by the 
action of the fermionic currents. The OPEs involving $n^l_{J,q}(z)$ are 
\beqa
&&J_E (z) n^l _{J,q}(w)
= \frac{(J-1/2)-l}{z-w}n^{l+1} _{J,q}(w), \n
&&J_F (z) n^l _{J,q}(w)
= \frac{(J-1/2)+l}{z-w}n^{l-1} _{J,q}(w), \n
&&J_H (z) n^l _{J,q}(w)
= \frac{2l}{z-w}n^{l} _{J,q}(w), \n
&&J_{H^{\prime}} (z) n^l _{J,q}(w)
=- \frac{1+2q}{z-w}n^{l} _{J,q}(w), \n
&&j_e (z) n^l _{J,q}(w)=\frac{-1}{z-w}
\left(\frac{J-q}{2J}N^{l+1/2} _{J,q}(w)
-\frac{(J-1/2)-l}{2J}{\cal N}^{l+1/2} _{J,q}(w) \right), \n
&&j_{\bar e} (z) n^l _{J,q}(w)=0, \n
&&j_f (z) n^l _{J,q}(w)=0, \n
&&j_{\bar f }(z)n^l _{J,q}(w)
=\frac{-1}{z-w}\left(\frac{J-q}{2J}N^{l-1/2} _{J,q}(w)
+\frac{(J-1/2)+l}{2J}{\cal N}^{l-1/2} _{J,q}(w) \right),  
\eeqa

\no and   
\beqa
&&J_E (z) {\bar n}^l _{J,q}(w)
= \frac{(J-1/2)-l}{z-w}{\bar n}^{l+1} _{J,q}(w), \n
&&J_F (z) {\bar n}^l _{J,q}(w)
= \frac{(J-1/2)+l}{z-w}{\bar n}^{l-1} _{J,q}(w), \n
&&J_H (z) {\bar n}^l _{J,q}(w)
= \frac{2l}{z-w}{\bar n}^{l} _{l,q}(w), \n
&&J_{H^{\prime}} (z) {\bar n}^l _{J,q}(w)
=\frac{1+2q}{z-w}{\bar n}^{l} _{J,q}(w), \n
&&j_e (z) {\bar n}^l _{J,q}(w)=0, \n
&&j_{\bar e} (z) {\bar n}^l _{J,q}(w)
=\frac{-1}{z-w}\left(\frac{J+q}{2J}N^{l+1/2} _{J,q}(w)
+\frac{(J-1/2)-l}{2J}{\cal N}^{l+1/2} _{J,q}(w) \right), \n
&&j_f (z) {\bar n}^l _{J,q}(w)
=\frac{-1}{z-w}\left(\frac{J+q}{2J}N^{l-1/2} _{J,q}(w)
-\frac{(J-1/2)+l}{2J}{\cal N}^{l-1/2} _{J,q}(w) \right),\n 
&&j_{\bar f }(z){\bar n}^l _{J,q}(w)=0.
\eeqa

\no Finally, the OPEs of the currents with  ${\cal N}^n _{J,q}(z)$ are
\beqa
&&J_E (z) {\cal N}^n _{J,q}(w)
= \frac{(J-1)-n}{z-w}{\cal N}^{n+1} _{J,q}(w), \n
&&J_F (z) {\cal N}^n _{J,q}(w)
= \frac{(p-1)+n}{z-w}{\cal N}^{n-1} _{J,q}(w), \n
&&J_H (z) {\cal N}^n _{J,q}(w)
= \frac{2n}{z-w}{\cal N}^{n} _{J,q}(w), \n
&&J_{H^{\prime}} (z) {\cal N}^n _{p,q}(w)
= \frac{2q}{z-w}{\cal N}^{n} _{J,q}(w), \n
&&j_e (z) {\cal N}^n _{J,q}(w)
=-\frac{J-q}{z-w}{\bar n}^{n+1/2} _{J,q}(w), \n
&&j_{\bar e} (z) {\cal N}^n _{J,q}(w)
=\frac{J+q}{z-w}n^{n+1/2} _{J,q}(w),\n
&&j_f (z) {\cal N}^n _{p,q}(w)
=\frac{J+q}{z-w}n^{n-1/2} _{J,q}(w), \n 
&&j_{\bar f }(z) {\cal N}^n _{p,q}(w)
= -\frac{J-q}{z-w}{\bar n}^{n-1/2} _{J,q}(w).
\eeqa

We would like to make a remark on the special case 
when $J=0, q\not= 0$. In this case the representation with the highest 
weight state (\ref{state-nst}) is typical. However, this representation
is infinite-dimensional, as is seen from the following bases of
the representation:
\beqa
&&N^m _{q, \pm}(z)=\left(\gm ^{-m}(z)
\mp \frac{m}{2} \gm ^{-m-1}(z) \ps (z)\psb (z) \right)V_{0,q}(z), \n 
&&~~~~~~~~~~~~~m=0,~-1,~-2,\cdots, \n
&&n^l _{q}(z)=\gm ^{(-l-1/2)} (z)\ps (z)V_{0,q}(z), 
~~l=-1/2,-3/2, \cdots, \n
&&{\bar n}^l _{q}(z)=\gm ^{(-l-1/2)} (z) \psb (z)V_{0,q}(z), 
~~l=-1/2,-3/2, \cdots, \n
&&{\cal N}^n _{q}(z)=\gm ^{(-nJ-1)}\ps (z) \psb (z) V_{0,q}(z), 
~~n=-1,-2,\cdots.
\eeqa

\subsection{$osp(2|2)^{(2)}$ primary fields}

The highest weight state of the algebra is 
\beq
V_{J,\pm}(z)={\rm exp} \{\frac{2}{\al _+}J(i \ph (z) \pm \ph ^{\prime}(z))\}
\eeq

\no where $J$ is any given complex number characterizing the
representation. The conformal dimension of the field is 
\beq
\Delta_{J,\pm}=0.
\eeq
So there are no typical representations and all representations are atypical.
It turns out that for the representation to be finite-dimensional, 
$J$ has to be an integer or
half integer. The full bases of the representation are
\beqa
&&\cT^m _{J,\pm}(z)=\left[
\left(-\gm (z)\right)^{J-m} 
\mp (J-m) \left(-\gm (z)\right)^{J-m-1}\ps (z)\psp (z) 
\right]V_{J,\pm}(z), \n 
&&~~~~~~~~~~~~~m=J,~J-1,\cdots, -(J-1),-J, \n
&&t^l _{J,\pm}(z)=\left(-\gm (z)\right)^{(J-1/2)-l}
\left(\ps (z) \mp \psp (z)\right) V_{J,q}(z),\n 
&&~~~~~~~~~~~~~l=(J-1/2),\cdots,-(J-1/2).
\eeqa 
$\cT^m _{J,\pm}(z)$ and $t^l _{J,\pm}(z)$ have $(2J+1)$ and $2J$ independent
components, respectively. So the dimension of the representation is
$4J+1$.

The OPEs of the currents with  $\cT^m _{J,\pm}(z)$ are
\beqa
&&J_E (z) \cT^m _{J,\pm}(w)= \frac{J-m}{z-w}\cT^{m+1} _{J,\pm}(w), \n
&&J_F (z) \cT^m _{J,\pm}(w)= \frac{J+m}{z-w}\cT^{m-1} _{J,\pm}(w), \n
&&J_H (z) \cT^m _{J,\pm}(w)= \frac{2m}{z-w}\cT^{m} _{J,\pm}(w), \n
&&J_{H^{\prime}} (z) \cT^m _{J,\pm}(w)= \frac{\mp 2J}{z-w}
  \cT^{m} _{J,\pm}(w), \n
&&j_e (z) \cT^m _{J,\pm}(w)=-\frac{J-m}{z-w}
t^{m+1/2} _{J,q\pm}(w), \n
&&j_{e^{\prime}} (z) \cT^m _{J,\pm}(w)=\pm \frac{J-m}{z-w}
t^{m+1/2} _{J,\pm}(w), \n
&&j_f (z) \cT^m _{J,\pm}(w)=-\frac{J+m}{z-w}
{t}^{m-1/2} _{J,\pm}(w), \n
&&j_{f^{\prime}} (z) \cT^m _{J,\pm}(w)=\pm \frac{J+m}{z-w}
{t}^{m-1/2} _{J,\pm}(w) .
\eeqa

\no We see that ${t}^{l} _{J,\pm}(z)$ 
are generated from $\cT^m _{J,\pm}(z)$ by the action of the 
fermionic currents. The OPEs involving $t^l_{J,\pm}(z)$ are 
\beqa
&&J_E (z) t^l _{J,\pm}(w)= \frac{(J-1/2)-l}{z-w}t^{l+1} _{J,\pm}(w), \n
&&J_F (z) t^l _{J,\pm}(w)= \frac{(J-1/2)+l}{z-w}t^{l-1} _{J,\pm}(w), \n
&&J_H (z) t^l _{J,\pm}(w)= \frac{2l}{z-w}t^{l} _{l,\pm}(w), \n
&&J_{H^{\prime}} (z) t^l _{J,\pm}(w)=\mp \frac{1+2J}{z-w}t^{l} _{J,\pm}(w),\n
&&j_e (z) t^l _{J,\pm}(w)=\frac{-1}{z-w}\cT^{l+1/2} _{J,\pm}(w), \n
&&j_{e^{\prime}}(z)t^l _{J,\pm}(w)=\frac{\pm 1}{z-w}\cT^{l+1/2} _{J,\pm}(w),\n
&&j_f (z) t^l _{J,\pm}(w)=\frac{1}{z-w}\cT^{l-1/2} _{J,\pm}(w),, \n
&&j_{f^{\prime}}(z)t^l _{J,q}(w)=\frac{\pm 1}{z-w}\cT^{l-1/2} _{J,\pm}(w). 
\eeqa

\section{Conclusions}

We have studied the conformal field theories associated with the current
superalgebras $osp(2|2)^{(1)}$ and $osp(2|2)^{(2)}$. We construct the
free field representations and screen currents of these two
superalgebras at general level $k$. We also construct the primary fields
corresponding to both typical and atypical representations. Both
conformal field theories have vanishing central charges. In the case of
$osp(2|2)^{(1)}$, there exists an infinite family of negative
dimensional primary operators so that the corresponding conformal field
theory is non-unitary. For the case of $osp(2|2)^{(2)}$, the dimension
of all primary fields vanishes and so they all correspond to
atypical representations of the current superalgebra. 
Our results provide a useful algebraic background in the study of
disordered systems using the supersymmetric method, which will be
investigated elsewhere.

\vskip.3in

\no {\bf Acknowledgments:}

This work is financially supported by Australian Research Council. 
One of the authors (Ding) is also supported partly by the 
Natural Science Foundation of China and a Fund from AMSS.

\bebb{99}

\bbit{Dot83}
V. Dotsenko and Vl. Dotsenko, {\it Adv. Phys.} {\bf 32}, (1983) 129.

\bbit{Lud90}
A. Ludwig, \npb {330} {1990} 639.

\bbit{Sha87}
R. Shankar, \prl {58} {1987} 2466.

\bbit{Car82}
J.L. Cardy and S. Ostlund, {\it Phys. Rev.} {\bf B25}, (1982) 6899.

\bbit{Lud94}
A. Ludwig, M. Fisher, R. Shankar and G. Grinstein, {\it Phys. Rev.} {\bf
B50}, (1994) 7526.

\bbit{Zir94}
M. Zirnbauer, {\it Ann. Physik} {\bf 3}, (1994) 513.

\bbit{Cau96}
J.-S. Caux, I.I. Kogan and A.M. Tsvelik, \npb {466} {1996} 444.

\bbit{Bas00}
Z.S. Bassi and A. LeClair, \npb {578} {2000} 577.

\bbit{Efe83}
K. Efetov, {\it Adv. Phys.} {\bf 32}, (1983) 53. 

\bbit{Mud96}
C. Mudry, C. Chamon and X.-G. Wen, \npb {466} {1996} 383.

\bbit{Ber95}
D. Bernard, e-print hep-th/9509137.

\bbit{Maa97}
Z. Maassarani and D. Serban, \npb {489} {1997} {603}.

\bbit{Gad91}
R. Gade and F. Wegner, \npb {360} {1991} {213}.

\bbit{Gur99}
S. Guruswamy, A. LeClair and A.W.W. Ludwig, e-print cond-mat/9909143.

\bibitem{BOo}
M. Bershsdsky and H. Ooguri, \plb {229} {1989} {374}.

\bbit{Ito}
K. Ito, \plb {252}{1990}{69}

\bbit{Bow96}
P. Bowcock, R-L.K. Koktava and A. Taormina, \plb {388} {1996} {303}.

\bbit{Ras98}
J. Rasmussen, \npb {510} {1998} {688}.

\bibitem{Wak}
M. Wakimoto, {\it Commun. Math. Phys. {\bf 104}}, (1986)605.



\bbit{FF3}
 B. Feigin and E. Frenkel, \cmp {128}{1990}{161}. 



\bbit{BMP3}
P. Bouwknegt, J. McCarthy and K. Pilch, {\it Prog. Phys. Suppl.} 
{\bf 102}, (1990) 67.

\bbit{Kur}
G. Kuroki, \cmp {142}{1991}{511}.

\bbit{Fren}
E. Frenkel, {\it Free Field Realizations in Representation Theory and 
Conformal Field Theory}, hep-th/9408109.




\bbit{Sha98}
A. Shafiekhani and W.S. Chung, Mod. Phys. Lett. {\bf A13}, (1998) 47.

\bbit{Lud00}
A.W.W. Ludwig, cond-mat/0012189.

\bbit{Dob93}
V.K. Dobrev, in {\it ``Espoo 1986, Proceedings, Topological and
Geometrical Methods in Field Theory"}, pp.93.

\bbit{Kob94}
K. Kobayashi, Z. Phys. {\bf C61}, (1994) 105.

\bbit{Bow97}
P. Bowcock and A. Taormina, Commun. Math. Phys. {\bf 185}, (1997) 467.

\bbit{Sem97}
A.M. Semikhatov, Theor. Math. Phys. {\bf 112}, (1997) 949.

\bibitem{Kac}
V. G. Kac, {\it Infinite Dimensional Lie Algebras}, third ed.,
Cambridge University press, Cambridge 1990.




\bbit{Din01}
X.M. Ding, M.D. Gould and Y.Z. Zhang, \plb {523} {2001} {367}.

\eebb

\end{document}